\documentclass[%
 reprint,
 amsmath,amssymb,
 aps,
]{revtex4-2} 
\pdfoutput=1
\usepackage{graphicx}
\usepackage{dcolumn}
\usepackage{bm}
\usepackage{braket}
\usepackage{textcomp}
\begin{document}

\newcommand{\fzeroone}{$\Lambda_{01}$}
\newcommand{\fonetwo}{$\Lambda_{12}$}
\newcommand{\fzerotwo}{$\Lambda_{02/2}$}
\newcommand{\hamil}[2][term]{\frac{\Omega_{#1#2}}{2}\hat{#1}\hat{#2}}
\newcommand{\zx}{$\hat{ZX}$}
\newcommand{\zy}{$\hat{ZY}$}
\newcommand{\zz}{$\hat{ZZ}$}
\newcommand{\ix}{$\hat{IX}$}
\newcommand{\iy}{$\hat{IY}$}
\newcommand{\iz}{$\hat{IZ}$}
\newcommand{\zi}{$\hat{ZI}$}

\preprint{APS/123-QED}

\title{Echo Cross Resonance gate error budgeting on a superconducting quantum processor}

\newcommand{\oqc}{
\affiliation{%
 Oxford Quantum Circuits, Thames Valley Science Park, 1 Collegiate Square, Reading RG2 9LH, UK\\
}%
}

\newcommand{\nqcc}{
\thanks{%
 Present Address:~National Quantum Computing Centre, Rutherford Appleton Laboratory,
Harwell Campus,
Didcot
OX11 0QX, UK\\
}%
}

\author{Travers Ward}

\author{Russell P. Rundle}

\author{Richard Bounds}

\author{Norbert Deak}

\author{Gavin Dold}

\author{Apoorva Hegde}

\author{William Howard}

\author{Ailsa Keyser}

\author{George B. Long}

\author{Benjamin Rogers}

\author{Jonathan J. Burnett}
\nqcc{}

\author{Bryn A. Bell}
\oqc{}

\date{\today}

\begin{abstract}
High fidelity quantum operations are key to enabling fault-tolerant quantum computation. Superconducting quantum processors have demonstrated high-fidelity operations, but on larger devices there is commonly a broad distribution of qualities, with  the low-performing tail affecting near-term performance and applications. Here we present an error budgeting procedure for the native two-qubit operation on a 32-qubit superconducting-qubit-based quantum computer, the OQC Toshiko gen-1 system. We estimate the prevalence of different forms of error such as coherent error and control qubit leakage, then apply error suppression strategies based on the most significant sources of error, making use of pulse-shaping and additional compensating gates. These techniques require no additional hardware overhead and little additional calibration, making them suitable for routine adoption. An average reduction of 3.7x in error rate for two qubit operations is shown across a chain of 16 qubits, with the median error rate improving from 4.6$\%$ to 1.2$\%$ as measured by interleaved randomized benchmarking. The largest improvements are seen on previously under-performing qubit pairs, demonstrating the importance of practical error suppression in reducing the low-performing tail of gate qualities and achieving consistently good performance across a device.
\end{abstract}

\maketitle

\section{\label{sec:Intro}Introduction}

Quantum computation at utility scale requires large numbers of qubits along with high-quality logical operations. Superconducting qubits are a leading approach, having demonstrated scaling to hundreds of qubits per device~\cite{willow, gao2024, AbuGhanem_2025} along with fast quantum logic gates~\cite{Fasciati2025, fast_1q, Spring2024}. In particular, fixed-frequency transmon qubits with capacitive couplers between neighbouring qubits have shown long coherence times and are an appealingly straightforward architecture for scaling, avoiding the additional complexity and noise of flux-tuneable elements~\cite{Wang2022-po}. The coaxmon architecture makes use of out-of-plane coaxial lines for qubit control and readout, avoiding on-chip control wiring, which leads to a tileable design with very low control crosstalk~\cite{Rahamim2017, Spring2022}. 

High-fidelity logic gates have been demonstrated with fixed-frequency superconducting qubits, but the speed and quality of two-qubit entangling gates (most commonly using the cross-resonance interaction~\cite{Corcoles2013}) is strongly dependent on the frequency detuning between neighbouring qubits~\cite{Hyppa2024}. Background interactions between coupled qubits can also become a significant error source, which is again dependent on their frequency detuning. Qubit frequency is difficult to precisely control in fabrication, leading to variability in gate quality across a quantum processing unit (QPU)~\cite{Morvan2022, Pappas_2024, kennedy2025}.
The presence of two-level systems (TLSs) also negatively impact gate operation, causing fluctuating coherence times and frequency instability \cite{Burnett2019, Schl_r_2019}. Hence typical gate fidelities across larger QPUs have tended to be lower than what can be achieved in isolated pairs of qubits. The lower end of the distribution of gate qualities across a larger scale device hampers utility for today's quantum applications and the ability to apply quantum-error correction~\cite{Mohseni2024}. This motivates the study of the error-budget across medium sized QPUs, along with targeted use of error suppression strategies to mitigate the largest sources of error, both in order to achieve the best performance and to inform future design requirements for scalability.

\begin{figure*}
    \centering
    \includegraphics[width=0.8\linewidth]{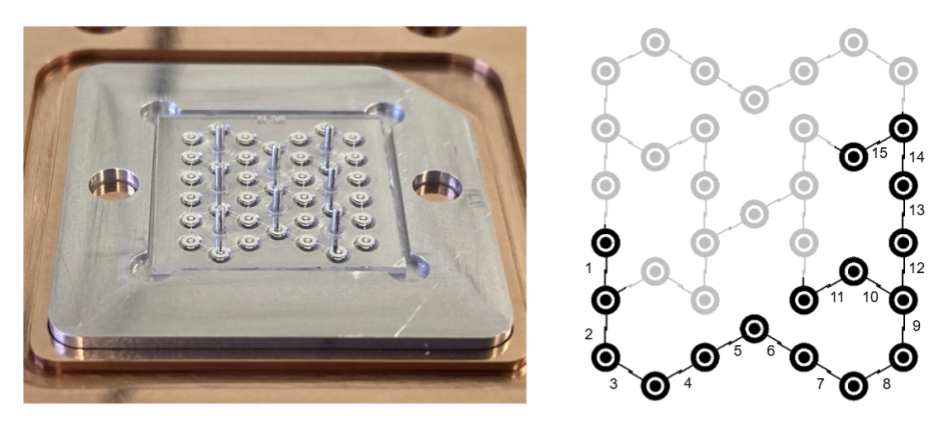}
    \caption{(Left) Photograph of a Toshiko QPU chip in an open sample package. 35 coaxmon qubits and 8 through-sapphire metal pillars are visible. (Right) Illustration of the qubit connectivity for Toshiko Tokyo-1. The pairs of qubits that were benchmarked are numbered for reference in later figures, while unused qubits and connections are shown grayed out.}
    \label{fig:toshiko}
\end{figure*}

In this paper, we present a detailed study of the errors affecting the echo cross resonance (ECR) gate~\cite{Sheldon_2016}, the native two-qubit operation, on an OQC Toshiko Gen-1 device. Through experimental measurements, we quantify three main sources of error: 
\begin{itemize}
    \item incoherent errors due to intrinsic qubit energy relaxation and dephasing,
    \item control qubit leakage errors, where the strong cross-resonant drive unintentionally moves the control qubit between computational states,
    \item coherent error, predominantly caused by the background ZZ interaction between the control and target qubit.
\end{itemize}
We then take a variety of gate-level error-suppression steps and find we can improve the overall performance of the device, focusing on a connected chain of 16 qubits. Across these qubits the median ECR gate error is improved from 4.6$\%$ with the native implementation to 1.2$\%$ with error-suppression, a 3.7x improvement, demonstrating that most of the coherent error and leakage can be removed using known techniques. What remains is the error originating from decoherence, and some unexplained error, for which we briefly discuss possible origins.

The layout of the paper is as follows. In Section~\ref{sec:system} we give a brief overview of OQC Toshiko Gen-1, describing the QPU and its characteristics. Section~\ref{sec:incoherent_error} discusses and quantifies errors arising from decoherence. In Section~\ref{sec:control_leakage} we measure the control leakage across the qubit pairs using an error amplification experiment, and then show that it can be well-suppressed by pulse-shaping. In Section~\ref{sec:coherent_error} we discuss sources of coherent error, and find that the largest errors can be suppressed by compensating rotations. Finally, in Section~\ref{sec:error_budget}, we summarize the overall error budget, before and after error-suppression, and compare it to the total error rates measured using interleaved randomized benchmarking.

\section{\label{sec:system}System Overview}

\begin{table}[h]
\centering
\begin{tabular}{l|r}
 Parameter & Value \\\hline
Qubit frequency (min/max, GHz) & 4.24 / 4.53 \\
Qubit anharmonicity (typ, MHz) & -182 \\
Readout frequency (min/max, GHz)~~ & ~9.63 / 10.27\\
$T_1$ median ({\textmu}s) & 69\\
$T_{2e}$ median ({\textmu}s) & 103 \\
$J$ coupling rate (typical, MHz) & 2.7 \\
Readout fidelity (median, $\%$) & 96 \\
$SX$ gate fidelity (median, $\%$) & 99.9
\end{tabular}
\caption{\label{tab:params1} Representative QPU parameters: the range of qubit frequency; typical anharmonicity $\alpha$; range of readout frequency; the excited state lifetime T1 and Hahn echo lifetime T2e; the qubit-qubit coupling rate $J$; the median readout fidelity; and the median fidelity of a one-qubit $SX$ (or $\sqrt{X}$) gate.}
\end{table}

OQC Toshiko is a quantum computing platform with a core QPU consisting of 35 fixed-frequency coaxmon qubits, of which 32 are connected and available. Fig.~\ref{fig:toshiko} shows a photo of a Toshiko chip mounted in its packaging, along with a diagram of the coupling graph. Toshiko systems are compatible with modern data centers, and have been integrated into co-location data-centers in London and Tokyo, with another integrated into a hybrid HPC/Quantum system at the Galicia Supercomputing Center (CESGA)~\cite{Cacheiro2025}. The results in this work were obtained with Toshiko Gen-1 (Tokyo), operating in a commercial data center environment.

The coaxmon architecture consists of transmon qubits constructed from concentric islands of superconducting material on one side of a substrate, with corresponding readout resonators defined on the reverse of the substrate - in this case the superconducting material is aluminium and the substrate sapphire~\cite{Acharya2025}. All control wiring is removed from the plane of the chip in a 3D architecture - coaxial control lines are anchored directly to the top and bottom of the QPU package, for qubit control and readout respectively, and terminate in pins aligned with each qubit and resonator~\cite{Rahamim2017}. Coupling arms extending from the outer island of each qubit are used to create a capacitive interaction between neighbouring qubits according to the coupling graph. Through-sapphire pillars are used to inductively shunt the chip enclosure, mitigating the effect of low-frequency cavity modes which would otherwise couple to the qubits~\cite{Spring2022} - the fabrication and machining process is described in Ref.~\cite{Acharya2025}. Representative device parameters are given in Table~\ref{tab:params1}.

\section{Incoherent Error\label{sec:incoherent_error}}
Decoherence mechanisms form fundamental limits to the time that a qubit can store a quantum state. Broadly this includes energy relaxation to the ground state, pure dephasing, and fluctuations in qubit frequency, where the relaxation and dephasing rates will themselves fluctuate over time due to mechanisms such as the interactions with unstable two-level-system (TLS) defects~\cite{Burnett2019}.

The durations of the ECR gates vary between particular pairs of qubits, ranging from 250ns to 460ns, during which time there is a small but non-negligible chance of an incoherent error occurring. This tends to be viewed as an intrinsic property of the qubits and a lower limit on the error-per-gate (EPG). To estimate the rate of incoherent errors, we make use of the measured $T_1$ and $T_{2e}$ coherence times of each qubit. From these it is possible to extract a relaxation rate and a dephasing rate, which are input to an analytic formula for the resulting error rate~\cite{chen2018}.

The coherence measurements used to calculate this error rate are repeated over many hours and then median values are taken, to average over fluctuations in coherence time. We find incoherent error probabilities of between $0.3\%$ and $0.8\%$ over the different qubit pairs and therefore different ECR gates considered. We note that this method uses medians from repeated measurements as being representative over the time fluctuations - this value could differ from the true value at the time the ECR gate is benchmarked, resulting in a discrepancy. We also do not account for the error induced by temporal fluctuations in qubit frequency, which could be a source of unexplained error. The ECR gate uses an echo sequence on the control qubit, which will make it robust to fluctuations in the control qubit frequency over time scales longer than the gate duration, however it will be more sensitive to fluctuations in the target qubit frequency.

\section{Control Qubit Leakage}\label{sec:control_leakage}
The cross resonance interaction generates a rotation of the target qubit's state conditional on the state of a control qubit, which is achieved by driving the control qubit at the resonant frequency of the target~\cite{Patterson2019}. The state of the control qubit should ideally remain constant during the operation of the gate, but in practice it experiences strong off-resonant driving pulses and will be perturbed to some extent. Primarily this will shift the frequency of the control qubit due to the AC Stark effect - but so long as it retains its computational state this only results in an accumulated phase error which is well canceled by the echo sequence of the ECR gate. Leakage effects between computational states are more detrimental to the gate - due to the pulsed nature of the driving, there will be some non-zero frequency content present at resonant transition frequencies of the control qubit, which can drive it between states at some small rate per pulse.

\begin{figure*}[t]
    \centering
    \includegraphics[width=1\linewidth]{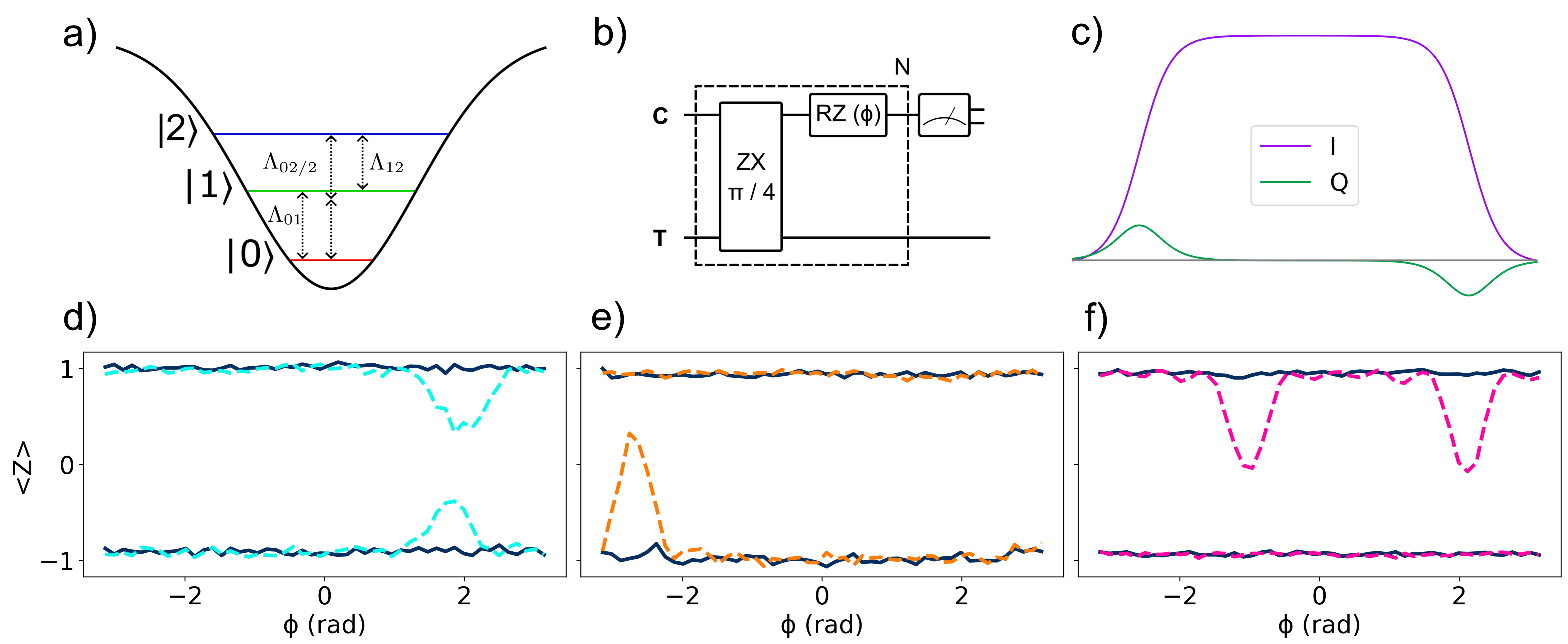}
    \caption{Control qubit leakage errors. (a) Illustrates the different leakage types occurring between energy levels. (b) Shows the circuit used to amplify and assess leakage errors, consisting of a repeat unit of a $ZX(\pi/4)$ pulse and a $Z(\phi)$ rotation. The angle $\phi$ is swept to locate an angle where the leakage errors add coherently, resulting in error amplification. (c) Leakage error suppression is achieved by modifying the pulse shape using DRAG. The original pulse shape (purple line) is supplemented by its derivative added in the quadrature $Q$ (green line), multiplied by some DRAG parameter. Representative results from the error amplification circuit are shown for pairs exhibiting (d) \fzeroone{}, (e) \fonetwo{}, and (f) \fzerotwo{} type leakage. The upper and lower lines correspond to the control qubit being initially prepared in $\ket{0}$ or $\ket{1}$ state, respectively. Dashed and solid lines show the results before and after leakage suppression.}
    \label{fig:leakage}
\end{figure*}

The control qubit transitions that cause the largest errors within our parameter regime are the single-photon transitions between $\ket{0}\leftrightarrow\ket{1}$ and $\ket{1}\leftrightarrow\ket{2}$ together with the two-photon transition between $\ket{0}\leftrightarrow\ket{2}$. We refer to these different types of control qubit leakage as \fzeroone{}, \fonetwo{} and \fzerotwo{} respectively, as illustrated in Fig.~\ref{fig:leakage}(a). These errors become prominent when the frequency of the cross-resonant drive tone approaches one of these transition frequencies, known as a frequency collision~\cite{Morvan2022}. By design, all control-target detunings should fall into the straddling regime (where efficient cross-resonant driving is possible) but avoid these collision conditions - however due to the uncertainty in fabrication and frequency crowding within the straddling regime, for some qubit pairs the detuning will fall close to a collision.

In order to assess the extent of each type of leakage error occurring on a particular qubit pair, we implement a simplified procedure as described in Ref.~\cite{Li2024}. In this, we form the error amplification circuit shown in Fig.~\ref{fig:leakage}(b), with repeated applications of the cross-resonance pulse generating a $ZX(\pi/4)$ rotation and a virtual $Z(\phi)$ rotation, where the phase $\phi$ is swept over $2\pi$. This is then repeated with the control qubit initially prepared in $\ket{0}$ or $\ket{1}$. Fig.~\ref{fig:leakage}(d, e, f) show examples of performing this procedure on different qubit pairs. We can identify the different forms of control leakage as follows:
\begin{itemize}
    \item \fzeroone{} when peaks occur at the same $\phi$ for both inital control states $\ket{0}$ and $\ket{1}$
    \item \fonetwo{} when a singular peak occurs for intial state $\ket{1}$ but the $\ket{0}$ state is unaffected
    \item \fzerotwo{} when two peaks occur for initial state $\ket{0}$ that are separated by $\pi$ radians - as a result of \fzerotwo{} being a two-photon transition, in this case the result is periodic in $\phi$ over $\pi$ instead of $2\pi$ radians.
\end{itemize}
The size of the peak at a given number of repetitions is used to estimate the leakage probability per $ZX(\pi/4)$ pulse. Using this technique, we find 
\begin{itemize}
    \item relatively large leakage errors for three of the fifteen qubit pairs considered, of $1-2\%$ per $ZX(\pi/4)$, of which two were \fonetwo{} and one \fzerotwo{} type,
    \item seven intermediate cases with $0.1 - 1\%$ leakage error,
    \item the five remaining cases had $<0.1\%$ leakage error.
\end{itemize}
Note that these are error rates per $ZX(\pi/4)$ rotation, and that an ECR gate contains two such rotations. These errors do not necessarily sum simply - the leakage from the two pulses could add constructively, or destructively cancel out. We attempted to directly measure the leakage per ECR gate in an error amplification circuit, but found that it was difficult to distinguish control leakage error from errors in the repeated application of $X$ gates. Hence when considering the ECR error budget we took the simple approximation of doubling the $ZX(\pi/4)$ leakage error, which is the average case over all all the phases with which the two contributions could add coherently.

\subsection{Leakage suppression}
Control qubit leakage can be minimised by modifying the pulse shape used to drive a $ZX(\pi/4)$ rotation, so as to remove spectral content that overlaps with a control qubit transition frequency. Derivative Reduction by Adiabatic Gate (DRAG) pulses provide a convenient method of defining pulses with minimal spectral content around particular frequencies~\cite{Motzoi2009}. Multi-derivative DRAG~\cite{Li2024} is able to suppress leakage errors over multiple transitions - here we find it is sufficient to use single-derivative DRAG to remove the single most significant leakage transition on a given pair, which reduces leakage errors to below $10^{-3}$, where it is less significant than incoherent error. This implies transforming the basic pulse shape as
\begin{equation}
    F(t) \rightarrow \left(1 + i\alpha\frac{d}{dt}\right)F(t)
\end{equation}
with $\alpha$ the DRAG parameter.

As discussed above, the leakage amplification measurement allows us to identify which types of leakage are occurring based on the number, location and symmetry of peaks. A DRAG parameter can then be calculated based on the detuning of the cross-resonance drive frequency $f_{\mathrm{CR}}$ from the affected transition, as
\begin{equation}
    \alpha = \frac{1}{2\pi (f_x - f_{CR})}
\end{equation}
when $f_x$ is the \fzeroone{} or \fonetwo{} transition frequency, or
\begin{equation}
    \alpha = \frac{1}{4\pi (f_x - f_{CR})}
\end{equation}
when $f_x$ is the \fzerotwo{}, i.e. it is smaller by a factor 2 for a two-photon transition. The leakage measurement is then repeated to verify the errors are suppressed, and $\alpha$ is fine-tuned by performing a local sweep.

In most cases this is sufficient to suppress leakage errors to $\sim10^{-4}$ or lower per $ZX(\pi/4)$ pulse. A few qubit pairs exhibit both \fzerotwo{} and one of the other leakage types. Due to the nature of a two-photon transition, the \fzerotwo{} error rate decreases quickly with the amplitude of the drive pulse, at a cost of increased pulse duration. For the cases with multiple leakage types, an increase of 10-20\% to pulse duration is sufficient to suppress \fzerotwo{}, and we then remove the remaining \fzeroone{} or \fonetwo{} leakage with DRAG. Although the increased duration will result in slightly more incoherent error per gate, we find the tradeoff beneficial in these cases.

\section{Coherent Errors\label{sec:coherent_error}}
An effective two-qubit Hamiltonian for the cross-resonance interaction can be written as
\begin{equation}
\begin{split}
    \hat{H}/\hbar = \frac{\Omega_{IX}}{2} \hat{IX}& + \frac{\Omega_{IY}}{2} \hat{IY} + \frac{\Omega_{IZ}}{2} \hat{IZ} + \frac{\Omega_{ZI}}{2} \hat{ZI} + \\ \frac{\Omega_{ZX}}{2} \hat{ZX}& + \frac{\Omega_{ZY}}{2} \hat{ZY} + \frac{\Omega_{ZZ}}{2} \hat{ZZ}
    \label{Eq:EffHamiltonian}
\end{split}
\end{equation}    
where $\Omega_{n,m}$ represents each term's rotation rate. This assumes that the interaction is block-diagonal in the control qubit, i.e. ignores control leakage as considered in the previous section.

Ideally all terms other than \zx{} would be zero, with the other contributions representing sources of coherent error. \zy{} is set to zero by correctly calibrating the phase of the cross-resonant drive, which moves contributions between the \zx{} and \zy{} terms. \ix{} and \iy{} are non-conditional driving terms on the target qubit - these are removed by calibration of a cancellation pulse, directly driving the target qubit, simultaneous with the cross-resonance pulse. The \zz{} term is the always-on interaction between coupled transmons, which creates a frequency shift conditional on the state of the other qubit. The \zi{} and \iz{} terms represent static frequency detunings of the control and target qubits respectively - but averaged over the state of the other qubit, which means that the presence of the \zz{} interaction also contributes to these terms. In addition, frequency calibration errors and any AC-Stark shift generated by the drive tone contribute to \zi{} and \iz{}~\cite{Corcoles2013}.

Note that the use of the echo in the ECR gate will naturally suppress some of the error terms. \ix{} and \zi{}, which commute with the dominant \zx{} term, are suppressed very effectively, whereas \iy{} and \zz{} are suppressed to first-order, but leave higher-order error contributions proportional to their commutator with \zx{}.

We characterize the unitary operation applied to the target qubit by our $ZX(\pi/4)$ operations using a subset of the measurements required for quantum process tomography, which is reduced from the full set using the assumption of unitarity. We find that the \iz{} term is the largest source of error we have identified - contributing up to $10\%$ EPG on qubit pairs with larger \zz{} interactions, with a mean EPG of $\sim2.3\%$ over the pairs considered. The \zz{} term itself is the next most significant form of coherent error, being suppressed to first-order by the ECR gate, contributing a maximum EPG of $\sim1.8\%$ and an average of $\sim0.4\%$. The other coherent error terms are well suppressed by the usual calibration and are not seen to contribute significantly.

\begin{figure}
    \centering
    \includegraphics[width=1\linewidth]{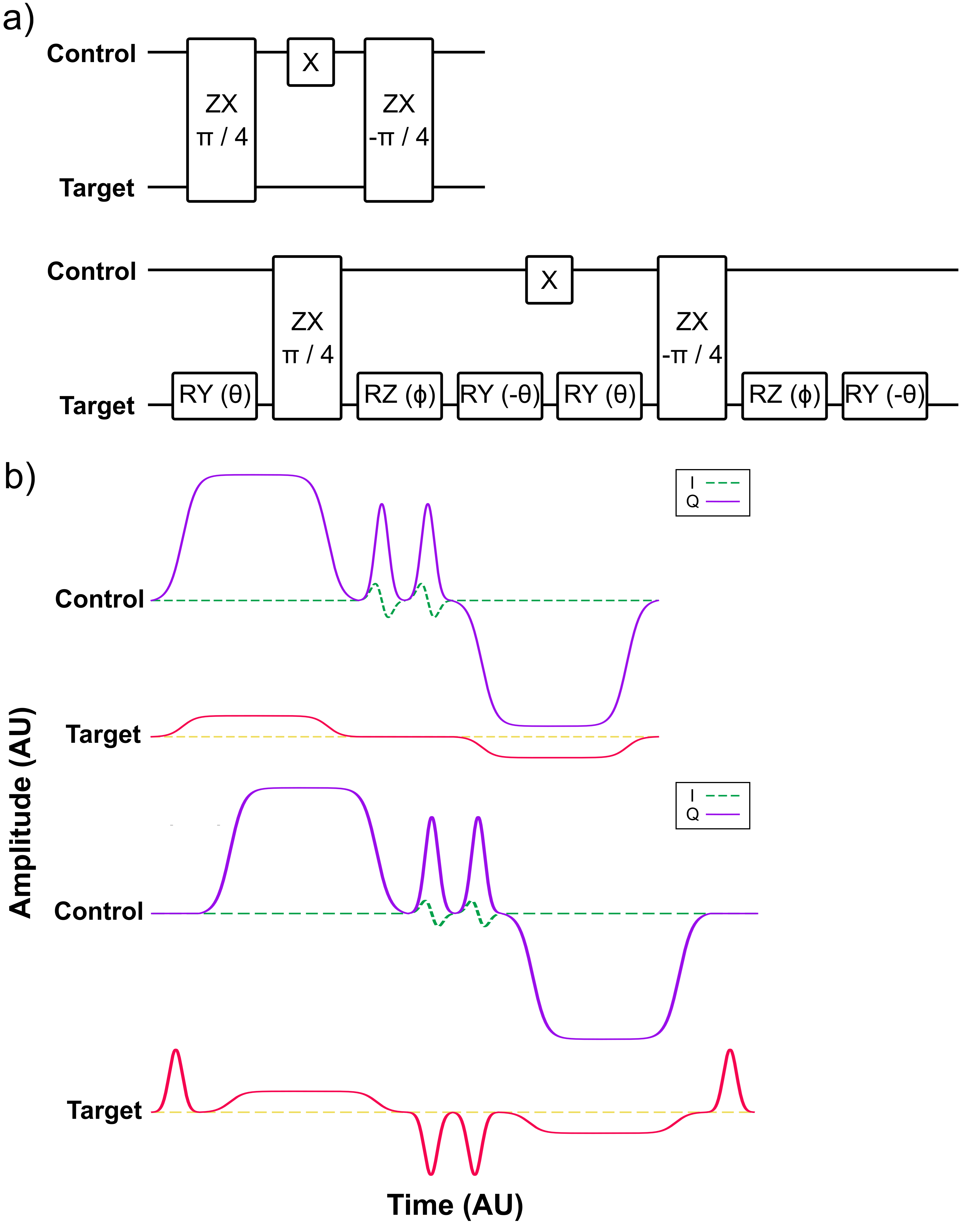}
    \caption{Qubit rotation and pulse level views of the ECR gate. a) - the naive implementation of the ECR, consisting of $ZX(\pi/4)$, $X$, and $ZX(-\pi/4)$ rotations. At the pulse level, the $X$ gate is implemented as two $SX$ rotations, each a Gaussian pulse with DRAG, which is the native one-qubit gate on Toshiko. b) - the ECR gate with corrections for coherent error, including $RZ$ and $RY$ rotations on the target qubit before and after each $ZX(\pi/4)$.}
    \label{fig:ecr_pulse_trains}
\end{figure}

\begin{figure*}
    \centering
    \includegraphics[width=0.9\linewidth]{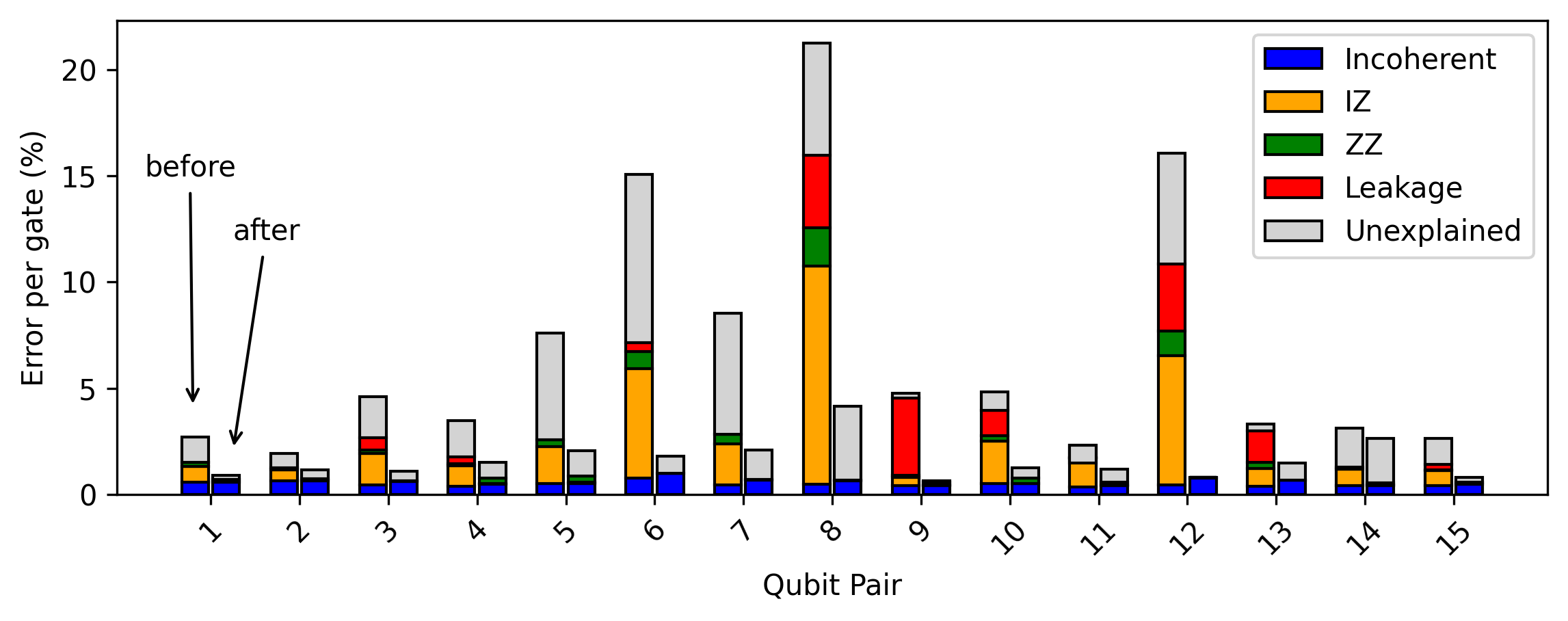}
    \caption{Error budget, showing the error-per-gate coming from incoherent errors, \iz{} and \zz{} coherent errors, and control leakage errors. For each qubit pair the errors before/after the application of error suppression are shown on the left/right.}
    \label{fig:after_budget}
\end{figure*}

\subsection{IZ Errors}
As discussed above, coherent \iz{} errors are on average the largest contributor to the EPG. They can be thought of as arising from a discrepancy between the target qubit's intrinsic frequency and the frequency it is driven at (both through cross-resonance and through its own drive pulses). In particular, the target qubit's frequency is calibrated with surrounding qubits in their ground state, and then when the two qubit gate is considered, the control qubit is also active and the target qubit's frequency is shifted due to the \zz{} interaction. From the effective Hamiltonian of Eq.~\ref{Eq:EffHamiltonian}, it would be correct to calibrate the target qubit's frequency to the mean of the values when the control is in $\ket{0}$ and $\ket{1}$ states, which would set $\Omega_{IZ}=0$. However this would imply modifying the target qubit's drive frequency depending on which two-qubit gate was being considered, which is incompatible with having a single calibration across the QPU, or with running multi-qubit circuits.

Instead we chose to correct on a per-gate basis by applying virtual-Z rotations to counteract the effect of \iz{} following each $ZX(\pi/4)$ gate. This is a simple and effectively `free' (in terms of time duration and error rate) method of compensation. It removes the \iz{} error from each gate considered in isolation within a single consistent calibration - however note that \iz{} terms would reappear if neighboring qubits other than the control were in an excited state, due to their \zz{} interaction with the target. To improve the performance of general quantum circuits, a rotary echo scheme may be beneficial to reduce interaction between target and spectators~\cite{Sundaresan2020}, while context-aware dynamical decoupling or circuit-compilation can help to mitigate background interactions between idling qubits~\cite{Seif2024, Coote2025}.

We choose the correction angle $\theta_c$ of the virtual-Z gate based on the experimentally measured effective Hamiltonian, so as to compensate the observed \iz{}. We note that this also transforms some of the \zx{} term into \zy{} - this can be seen from the Baker-Campbell-Hausdorff series for

\begin{equation}
     R_Z(\theta_c)~e^{i\hat{H}}= e^{-i\frac{\theta_c}{2}\hat{I} \hat{Z}} e^{i\hat{H}} = e^{i\hat{H}'}
\end{equation}
with $\hat{H}$ from Eq.~\ref{Eq:EffHamiltonian}, and $\hat{H}'$ the modified Hamiltonian including the effect of the correction.
\begin{equation}
    \hat{H'} = \hat{H} - \frac{\theta_c}{2}\hat{I} \hat{Z} + i\frac{\theta_c}{4}[\hat{H},\hat{I}\hat{Z}] + ...
\end{equation}
where assuming \zx{} is the dominant term in $H$, the first-order term $[\hat{H},\hat{I}\hat{Z}]$ will mostly consist of \zy{}. This is corrected by modifying the phase of the cross-resonance pulses.

Repeating the tomography procedure from Section~\ref{sec:coherent_error} after these corrections, we find the estimated errors from the \iz{} term in the effective Hamiltonian is reduced to $<10^{-3}$ EPG for all pairs.

\subsection{ZZ Suppression}

In most cases, this simple correction of \iz{} combined with the ECR gate's natural suppression of \zz{} is sufficient to suppress coherent errors to below the level of incoherent error. In 4 out of the 15 cases presented here however, the \zz{} interaction is large enough that it still generates significant errors in the ECR gate and requires further active suppression. These more extreme cases result from the control-target detuning approaching the qubits anharmonicity, where the \zz{} interaction becomes large~\cite{Zhao_2020} and \fonetwo{} leakage is also likely to be an issue. As with control leakage errors, these cases come about due to the need to balance efficient gates with avoiding collisions during QPU design, combined with the fabrication uncertainty in qubit frequency targeting.

The \zx{} and \zz{} operators can be transformed into one another via basis rotations about the \iy{} axis. Hence surrounding each $ZX(\pi/4)$ pulse with $R_Y(\theta)$ and $R_Y(-\theta)$, for some well-chosen $\theta$, should transform the unwanted \zz{} interaction into additional \zx{} driving. In practice we begin by adding a short $Y$ pulse on the target qubit before and after each $ZX(\pi/4)$, choosing the amplitude roughly such that
\begin{equation}
    \theta=\mathrm{arctan}\left(\frac{\Omega_{ZZ}}{\Omega_{ZX}}\right) \approx \frac{\Omega_{ZZ}}{\Omega_{ZX}},
\end{equation}
and then repeat our tomography procedure from Section~\ref{sec:coherent_error}, updating the amplitude so as to suppress the \zz{} term to zero. This also results in an increase of the \zx{} rotation to $\sqrt{\Omega^2_{ZX}+\Omega^2_{ZZ}}$, which is compensated for by decreasing the amplitude of the cross-resonance pulses.

Note that the \zz{} error was only removed on a select few pairs. This was due to the increased total gate duration from the additional pulses leading to a larger increase in coherent error than the original \zz{} error term. In the cases where this technique was used, we see good elimination of the \zz{} term in the gate tomography, and measure improved fidelities with interleaved randomized benchmarking.

\section{Error budget and comparison to Randomized Benchmarking}\label{sec:error_budget}

Finally we make use of interleaved randomized benchmarking (IRB) ~\cite{Mageson2012} to measure total EPG, then compare this to the individual sources of error. For each pair IRB was run, interleaving the ECR gate between randomly chosen two-qubit Clifford operations for 30 different choices of random sequence. This was repeated for both the naive implementation of the ECR gate and the error-suppressed version.

The overall error budget is shown in Fig~\ref{fig:after_budget}, containing the EPG estimates coming from different sources before and after error suppression. Any excess error on top of this, found by IRB, is included as 'unexplained' error. It can be seen that control leakage and coherent error are largely eliminated, with incoherent error the largest remaining source of error. We find that the randomized benchmarking EPGs closely follow the total error budget for each pair, but generally also contain significant unexplained error.

One possible source of excess error is the $X(\pi)$ rotation applied to the control qubit during the ECR gate, which we have not considered above under the assumption that the dominant errors were caused by the two-qubit $ZX(\pi/4)$ rotations. We note that in the context of the two-qubit gate, the $X(\pi)$ will be affected by coherent errors coming from the \zz{} interaction which could be missed when benchmarking the single-qubit gates in isolation. 

 Before error suppression, the median value of unexplained error is 1.7$\%$, and for the worst performing pairs it generally becomes more significant, with a maximum of 7.9$\%$. After error suppression, the unexplained excess error becomes comparable to or larger than the known errors - however, the median of the unexplained EPG is reduced to $0.6\%$. This could be interpreted as some systematic under-estimation of the error from known sources, such that the excess error is also reduced by the error suppression; alternatively, the excess error could originate from other sources that are also tangentially affected by the error suppression.

\begin{figure}
    \centering
    \includegraphics[width=1\linewidth]{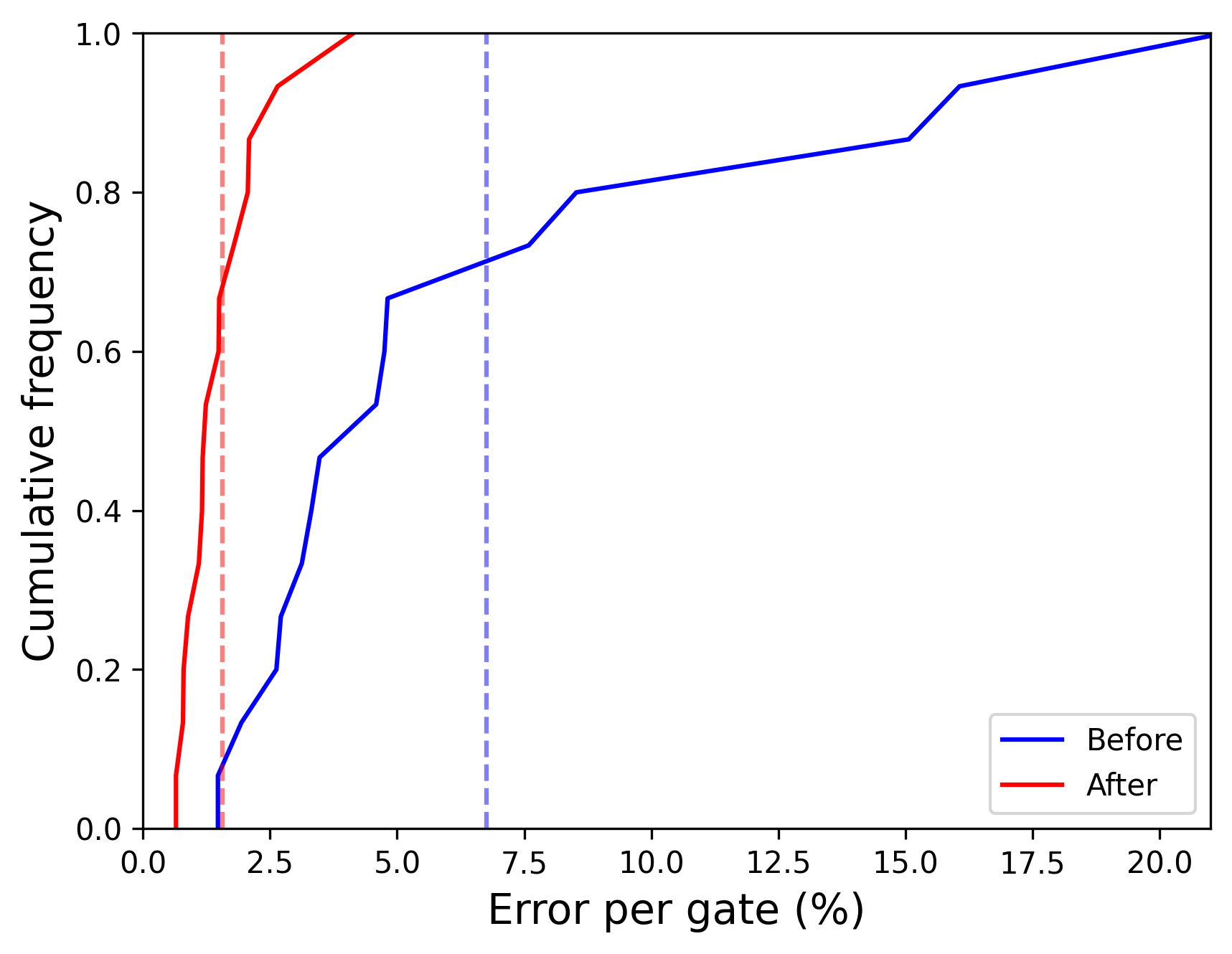}
    \caption{Comparison of interleaved benchmarking before and after error-suppression. Cumulative distributions of measured EPG over the separate qubit pairs. Dashed lines show the respective mean values.}
    \label{fig:before_vs_after}
\end{figure}

Figure.~\ref{fig:before_vs_after} shows the average fidelity from interleaved randomized benchmarking, before and after the error suppression. This shows the error rates have dramatically improved, with the median error per gate over the 16Q chain decreased from 4.6\% to 1.2\%, the mean improving from 6.75\% to 1.6\%, and the EPG of the best individual gate improving from 1.5\% to 0.6\%. However it is particularly the gates with the lowest fidelity before error suppression which show the largest improvements.

\section{Discussion}
In this work we have introduced a new error budgeting and calibration procedure on an OQC Toshiko Gen-1 device. First, we identified and measured control qubit leakage. We then suppressed this using a combination of single-derivative DRAG and modifications to the pulse duration and amplitude. This technique is extendable to deal with multiple leakage error transitions by using multi-derivative DRAG, where it has been shown to suppress leakage rates to $\sim 10^{-4}$~\cite{Li2024}.

Next, we tackled the coherent errors generated by \iz{} and \zz{} terms in the effective Hamiltonian of the gate. We found that the \iz{} error is the dominant source of errors in the ECR gates, and that the introduction of compensating virtual-Z gates is sufficient to suppress it. We also show how pre/post compensating pulses can be used to suppress the \zz{} error term. In the future, this \zz{} suppression may be improved by moving the compensating pulse to within the rise time of the cross-resonance pulse itself as shown in \cite{Li2024}, saving on gate time and making it viable for use on every qubit pair. Alternatively, the addition of a rotary echo scheme to the ECR gate could be used to suppress these terms as well as reducing spectator errors~\cite{Sundaresan2020}.

The discrepancy between the characterized errors and the IRB results indicates the presence of other forms of error that have not yet been identified or measured. One possible error source that we have not considered is the $\hat{X}(\pi)$ operation on the control qubit implementing the echo in the ECR gate. While we expect this operation to be relatively high fidelity compared to longer cross-resonance pulses, it could still be affected by coherent errors coming from miscalibration, leakage to the control qubit's $\ket{2}$ state, and it will be affected by the \zz{} interaction with the target qubit occurring during the gate duration. While calibrations can be improved and leakage can be addressed via DRAG and other pulse shaping techniques~\cite{Motzoi2009}, \zz{} interaction occurring over the time scale of a single qubit gate is potentially more challenging to suppress.

Overall, we observe a 3.7x reduction in error rates through the application of error suppression techniques, which introduce no additional hardware overhead and which can fairly easily be integrated into standard calibration routines. The improvement factor is comparable to that seen in Ref.~\cite{Li2024} - there the achieved error rates were lower, however here we have benchmarked a larger number of gates with wider variability.  In particular we see the biggest improvements for the qubit pairs that were previously performing poorly, reducing the tail of low-fidelity gates. This will lead to greater utility for running quantum circuits and algorithms, where performance is often limited by a 'weak-link' in a chain of connected qubits.

Beyond improving the present devices, error budgeting can also help to inform future design choices, for example in understanding the dominant errors and their dependence on frequency detunings, particularly in the context of post-fabrication frequency tuning~\cite{kennedy2025}. The fact that known error suppression techniques can reduce other error sources to below the level of incoherent errors also suggests that improving gate speed relative to the coherence time is required for further improvement, and that quantum gate schemes and QPU parameters like frequency detuning and qubit-qubit couplings strengths should be chosen with this in mind.

\begin{acknowledgments}
We thank Connor D. Shelly and Peter Leek for comments on the manuscript. We acknowledge the support of the Oxford Quantum Circuits Advanced Product Development team and the Material Science and Device Engineering team.
\end{acknowledgments}

\bibliographystyle{unsrt}
\bibliography{citations}

\end{document}